\documentclass[
    ,final            
  ]
  {aipproc}

\layoutstyle{6x9}

\newcommand{\msun}{\ifmmode\mbox{M}_{\odot}\else$\mbox{M}_{\odot}$\fi}
\newcommand{\lsun}{\ifmmode\mbox{L}_{\odot}\else$\mbox{L}_{\odot}$\fi}
\newcommand{\rsun}{\ifmmode\mbox{R}_{\odot}\else$\mbox{R}_{\odot}$\fi}

\begin{document}

\title{The Observed Spin Distributions of Millisecond Radio and X-ray
Pulsars}


\author{J.W.T. Hessels}{address={Astronomical Institute ``Anton
Pannekoek'', University of Amsterdam, Kruislaan 403, 1098 SJ
Amsterdam, The Netherlands; J.W.T.Hessels@uva.nl}}

\begin{abstract}
We consider the currently observed spin distributions of various types
of neutron stars, including isolated and binary radio millisecond
pulsars in the Galactic plane and globular cluster system as well as
neutron stars in low-mass X-ray binary systems where the spin rate is
known either through coherent pulsations or burst oscillations. We
find that the spin distributions of isolated and binary radio
millisecond pulsars are statistically different, at least for those
residing in globular clusters, with the binary pulsars being on
average faster spinning.  This result is likely to hold despite
observational biases still affecting the observed spin distribution.
A possible explanation for this is that the isolated radio millisecond
pulsars are on average older than those in binary systems.
\end{abstract}

\maketitle


\section{Introduction}

The number of known\footnote{The 61 known radio MSPs in the Galactic
plane are from the ANTF catalog
\citep[http://www.atnf.csiro.au/research/pulsar/psrcat]{mhth05}, where
we have only selected sources with $B_{surf} < 10^9$\,G (with only a
few exceptions, this is equivalent to selecting sources with
$\nu_{spin} > 50$\,Hz).  The 124 known radio MSPs in GCs with
$\nu_{spin} > 50$\,Hz are from the on-line catalog of P. Freire (see
http://www.naic.edu/$\sim$pfreire/GCpsr.html).  The accreting
millisecond pulsars are taken from \citet{wkbs08}, where we have
included all 10 known sources with accretion-powered pulsations as
well as the 8 sources where the spin rate has only been inferred
through the detection of burst oscillations (in at least two separate
bursts).} neutron stars (NSs) with millisecond spin periods has
roughly doubled in the last several years, providing a larger sample
with which to consider the spin rate distribution (Table~\ref{tab:a}).
The spin distributions of various sub-classes of NSs should be
intimately linked with the nature of the accretion processes in
low-mass X-ray binary systems (LMXBs) that spin up NSs (i.e. pulsar
``recycling'') and/or limit their spin-up, such as magnetic torque
braking \citep{gl78,aghw05,wz97} or gravitational wave emission
\citep{pp78,wag84,bil98}.  Thus, a lot of interesting physics is
likely to come from mapping these intrinsic distributions and by
comparing them to better understand the evolutionary link between NSs
LMXBs, presumably the progenitors\footnote{Though it may also be that
a significant fraction of the MSPs are formed in the accretion-induced
collapse of a white dwarf \citep[e.g.][]{fw07}.} of the radio millisecond pulsars
(MSPs), binary radio MSPs, and isolated MSPs (which are presumably
still {\it formed} in a binary system, but where the companion star
was somehow lost or destroyed).

\begin{table}
\begin{tabular}{llll}
\hline
Type            & Number & Period (ms) & Frequency (Hz) \\
                &        & mean/median & mean/median    \\
\hline \hline
All MSPs        & 185    & 5.0 / 4.4   & 252 / 227 \\
Binary MSPs     & 117    & 5.0 / 3.8   & 267 / 262 \\
Isolated MSPs   &  68    & 5.1 / 4.9   & 225 / 202 \\
\hline
All GC MSPs     & 124    & 4.8 / 4.2   & 258 / 239 \\
Bin. GC MSPs    &  73    & 4.5 / 3.6   & 288 / 278 \\
Iso. GC MSPs    &  51    & 5.3 / 5.0   & 215 / 198 \\
\hline
All Field MSPs  &  61    & 5.5 / 4.6   & 239 / 215 \\
Bin. Field MSPs &  44    & 5.8 / 4.6   & 233 / 217 \\
Iso. Field MSPs &  17    & 4.6 / 4.9   & 253 / 205 \\
\hline
VLMBPs ($f(m_1,m_2) < 10^{-4}$\,\msun) &  21    & 3.3 / 3.5   & 332 / 287 \\
LMBPs  ($f(m_1,m_2) > 10^{-4}$\,\msun) &  88    & 5.4 / 3.9   & 254 / 254 \\
Eclipsing       &  22    & 4.4 / 3.5   & 332 / 282 \\
Not Eclipsing   &  95    & 5.1 / 3.9   & 252 / 255 \\
$P_{orb} < 0.5$\,d &  37    & 4.5 / 3.6   & 288 / 274 \\
$P_{orb} > 0.5$\,d &  72    & 5.2 / 3.9   & 259 / 259 \\
\hline
All LMXBs       &  18    & 2.7 / 2.3   & 436 / 438 \\
AMXPs           &  10    & 3.3 / 2.6   & 367 / 389 \\
Burst Osc.      &   8    & 2.0 / 1.7   & 521 / 574 \\
\hline \hline
\end{tabular}
\caption{Average and median values of spin for various classes of NSs}
\label{tab:a}
\end{table}

\section{Discussion of Observed Populations}

\begin{figure}
\includegraphics[width=.6\textwidth,angle=270]{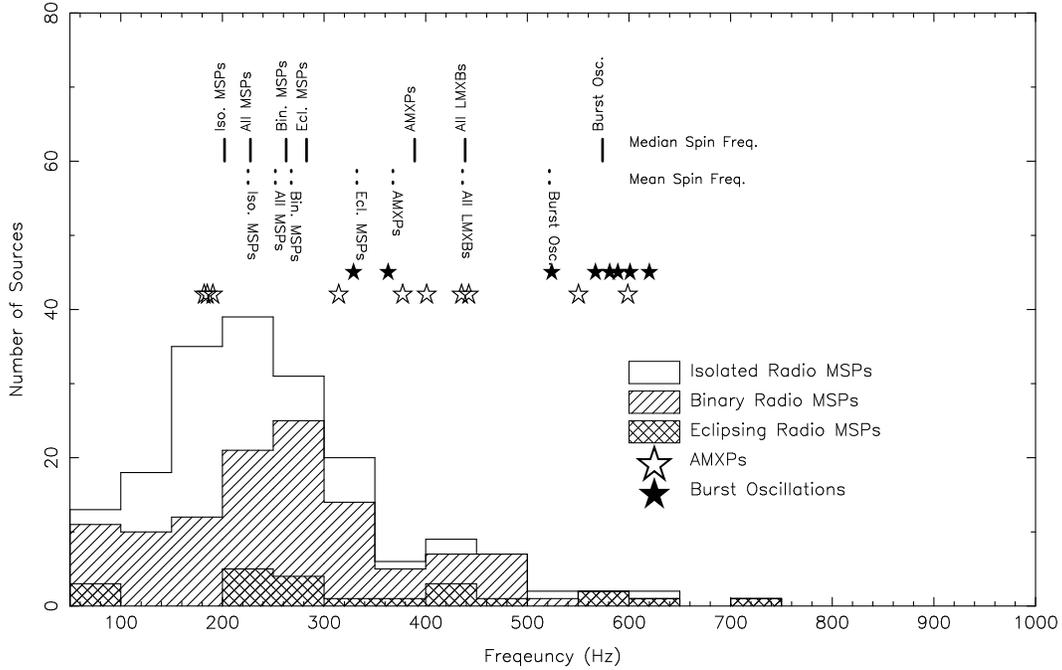}
\caption{The histogram shows the spin frequency distributions of all
known radio MSPs (isolated, binary, and eclipsing, including both those in the
field and in GCs).  The spin frequencies of the 10 known AMXPs and 8
known NMPs are shown above.  The average and median spin frequencies of
these NS sub-classes are also shown.} \label{fig:a}
\end{figure}

{\bf Radio MSPs}: In Figure~\ref{fig:a} we plot the spin distribution
of all known radio MSPs, both those in the field and in globular
cluster (GCs), as well as the accreting NSs in LMXBs where the spin
rate is known. The spin distribution of all radio MSPs combined shows
a peak around $\nu_{spin} = 200$\,Hz, with a sharp decline and tail at
higher spin frequencies out to the highest known spin frequency of
716\,Hz\citep{hrs+06}.  The form of the distribution at the highest
spin frequencies is likely still strongly affected by observational
biases against detecting the fastest-spinning radio pulsars (e.g., due
to scattering and binary motion, see also \citep{hrs+07} for a
detailed discussion of this), whereas the distribution below 200\,Hz
is more likely to be intrinsic. In Table~\ref{tab:a} we compare the
mean and median spin rates of different sub-classes of radio and X-ray
MSPs.  One of the simplest divisions one can make in the population of
radio MSPs is between those with a binary companion and those which
are isolated. We find statistical evidence that the spin distributions
of binary and isolated radio MSPs are in fact different.  A
Kolmogorov-Smirnov (KS) test of the two distributions indicates that
there is only a 0.03\% chance that the two observed populations are
derived from the same instrinsic distribution.  Comparing the average
and median spin frequencies, we also see that the binary radio MSPs
are on average faster-spinning than those that are
isolated\footnote{We also note \citep[see also][]{hrs+06}, that there
is some evidence (though the population of such objects is small) that
the fastest-spinning neutron stars are preferentially either in
eclipsing systems and/or have a very low mass companion \citep[the
so-called ``very low-mass binaries'', or VLMBs, see][]{fre05}.  This
could mean that a large fraction of the fastest-spinning radio MSPs
are self-obscuring and consequently very hard to detect.}. One expects
observational biases still present in the observed distribution to
make the binary pulsars appear to spin on average {\it less} rapidly
than isolated ones.  The fact that we observe the opposite scenario
implies that, on average, binary MSPs do indeed intrinsically spin
faster (note that the luminosity distributions of isolated and binary
MSPs, both those in the field and in GCs, appear to be
indistinguishable \citep{hrs+07,lmcs07}).

However, further sub-dividing the isolated and binary radio MSPs into
those found in the Galactic plane versus those found in GCs, we see
that there is (as yet) no statistically discernable difference in the
spin distributions of isolated and binary MSPs in the Galactic plane
\citep[as noted by][]{lmcs07}, whereas the spin distribution of
binaries and isolated MSPs in GCs are markedly different (KS test
indicates 0.0023\% chance that the two observed populations are drawn
from the same distribution).  The smaller known population of field
MSPs, coupled with observational biases, could be responsible for this
observed dichotomy.  We suspect that the spin distribution of binaries
in GCs is possibly less affected by observational bias than that in
the Galactic plane because repeated, targetted, deep searches
incorporating sophisticated algorithms for detecting binary signals
have been made.  Indeed, as Table~\ref{tab:a} shows, the median spin
frequencies of isolated pulsars in the field and in GCs are the same,
whereas the binaries in GCs have a much higher median spin frequency
than those in the field.  We think it is unlikely, though this issue
warrants further study and consideration, that this difference is due
to the dense stellar environments of GCs, which subject a resident NS
to potentially several interactions during its life in the cluster
core.  Thus, we predict that as the population of field MSPs
increases, it will become apparent that here too the binary MSPs are
on average spinning faster.  If this is not found to be the case, an
explanation relying on the properties of GC environments will have to
be found.

Given their overall similar spin rates, it appears likely that binary
and isolated MSPs are formed in a similar process.  It still remains
possible however that isolated and binary pulsars are formed in
somewhat different processes but achieve the same approximate final
spin rates because of a common spin limiting mechanism.  Here we
consider only the former, simpler scenario, and suggest an
evolutionary link between them.

A possible, simple explanation for the observed difference in spin rate
between the isolated and binary MSPs is that the isolated pulsars are
older than those in binaries and have consequently had more time to
spin down after they were recycled.  Another reason to favour this
explanation is that isolated MSPs presumably had binary companions
earlier in their lives but have lost or destroyed them over time
\citep[e.g.][]{rs86}.  Using the following expression for spin-down timescale
($\Delta t$) from initial spin $\nu_{\circ}$ to final spin $\nu_1$
(assuming moment of inertia $I = 10^{45}$\,g cm$^2$ and stellar radius
$R = 10^6$\,cm):

\begin{equation}
\frac{1}{\nu_1^2} - \frac{1}{\nu_{\circ}^2} = 6.15 \times 10^{-7}
\left(\frac{B}{10^8{\rm Gauss}}\right)^2 \left(\frac{R}{10^6{\rm
cm}}\right)^6 \left(\frac{\Delta{t}}{{\rm Gyr}}\right)
\end{equation}

\noindent we find that it takes 3\,Gyr for a pulsar with $B_{surf} =
2.5 \times 10^8$\,Gauss (the median $B_{surf}$ of MSPs in the
field\footnote{When considering physical parameters that are estimated
from spin-down, $\dot{P}$, we will restrict ourselves to those MSPs in
the field, who's spin-down, unlike in GCs, is not strongly
contaminated by an external gravitational field.}, where we assume
$B_{surf} \propto \sqrt{P\dot{P}}$) to spin down from the median spin
frequency of GC binaries (280\,Hz) to the median spin frequency of
isolated GC MSPs (200\,Hz).  This timescale can be accomodated by the
ages of MSPs, which have median characteristic ages\footnote{Note
however that \citet{fw07} argue that $\tau_c$ is 1.5 times the true
pulsar age in the case of MSPs} $\tau_c = 6$\,Gyr (defined as $\tau_c
\equiv P/2\dot{P}$).  We find only weak evidence that the isolated
MSPs in the field (median $\tau_c = 7.2$\,Gyr) are significantly older
than those in a binary system (median $\tau_c = 5.6$\,Gyr).  The two
distributions in $\tau_c$ are statistically consistent with each
other. Again, this could be due to low statistics.

An alternative explanation is that isolated MSPs have, on average,
larger magnetic fields than those in binary systems, and consequently
spin down on a shorter timescale.  Such high-$B_{surf}$ MSPs would
also potentially be more capable of ablating their companion star and
becoming isolated.  However, comparing the magnetic field distribution
of the isolated and binary pulsars in the field, there is as yet no
evidence that the distributions of $B_{surf}$ differ significantly.  
Again, we will likely have to wait until the known population of field
MSPs is significantly larger (or a larger number of GC MSPs have
reliable $\dot{P}$ estimates) before we can carefully test this
hypothesis.

{\bf NSs in LMXBs}: The population of still accreting NSs with known
spin rates is unfortunately still comparatively quite small.  However,
some interesting differences with the much larger known population of
radio MSPs are already apparent \citep[see also][for a much more
detailed discussion of this]{fw07,del08}.  For instance, the accreting
MSPs are on average {\it much} faster spinning than the radio MSPs
(Table~\ref{tab:a}). 
Although such a scenario is potentially over-simplistic
\citep[consider][where the link between LMXBs and MSPs is considered
from the viewpoint of binary evolution]{del08}, could the differences
in spin rate distributions simply demonstrate the often presumed
evolutionary trend from LMXB to binary MSP to isolated MSP, where
isolated MSPs are simply older than binary MSPs?  In other words, if
accretion were to suddenly shut off and these sources became radio
pulsars spinning at their current rate, would they naturally evolve
into the observed population of radio MSPs (igoring for the time being
orbital parameters)?  From Eqn. 1, the implied timescale is 2\,Gyr to
go from the 440-Hz median spin frequency of the accreting MSPs to the
280-Hz median spin frequency of the binary radio MSPs in GCs.  This is
not a completely unreasonable timescale, but may be too long to
accomodate certain young binary MSPs.  This could indicate that
further spin-down, in addition to normal magnetic dipole braking, may
be necessary.  We are planning on investigating this further. 


\begin{theacknowledgments}
J.W.T.H. is an NSERC post-doctoral fellow with a supplement from the CSA.
We thank Anna Watts, Scott Ransom, Paulo Freire, and Ingrid Stairs for useful
discussions and comments.
\end{theacknowledgments}


\end{document}